\documentclass[11pt]{article}

\usepackage[utf8]{inputenc}
\usepackage[T1]{fontenc}
\usepackage[margin=1in]{geometry}
\usepackage{amsmath}
\usepackage{amssymb}
\usepackage{amsfonts}
\usepackage{graphicx}
\usepackage{booktabs}
\usepackage{array}
\usepackage{caption}
\usepackage{float}
\usepackage{setspace}
\usepackage[hidelinks]{hyperref}

\graphicspath{{figures/}}

\allowdisplaybreaks

\begin{document}

% ============================================================
% Title
% ============================================================
\begin{center}
    {\Large\bfseries AI-Augmented Closed-Loop Quality Engineering: A Reference Architecture for Continuous Software Quality Intelligence\par}
\end{center}

\vspace{0.5em}

\begin{center}
    {\large Dimple Bajaj\par}
\end{center}

\vspace{1em}

\vspace{1em}

% ============================================================
% Abstract
% ============================================================
\noindent\textbf{Abstract}

The quality of software engineering is still under a challenge due to disjointed processes between requirements, testing, and production, which hinders the opportunity to implement quality strategies in consecutive releases. Existing approaches tend to be fixed-model or single-optimization approaches and lack production feedback learning mechanisms. The paper at hand proposes a closed-loop reference architecture of continuous software quality intelligence with AI enhancements. The model synthesizes requirement feature mining, risk-based test prioritization, defect prediction, and production incident analysis as an element of a feedback-based pipeline. A limited feedback learning model is introduced that is used to propagate the production signal-based on defect severity and incident impact- to the following release to ensure stability, and the time. The method is evaluated using a semi-synthetic test dataset of 4,500 requirements, 27,049 test cases, 13,089 defects and 7,841 incidents in six release cycles. The experimental results show that the proposed system reduces the defect leakage by 0.19 to 0.13, increases the effectiveness of the detection system to 0.72 to 0.84, and shortens the test execution by up to 35 percent compared to the non-adaptive baselines. The changes are stable release to release. The findings indicate that through the integration of feedback-based learning in a closed-loop architecture, it can be continued to enhance quality process, which offers practical foundation of adaptive quality engineering of software.

\vspace{0.5em}
\noindent\textbf{Keywords:} Closed-loop quality engineering, feedback-driven learning, defect prediction, test prioritization, software quality intelligence, production analytics.

% ============================================================
% 1. Introduction
% ============================================================
\section{Introduction}

The growing complexity of software systems and the need to get fast and iterative releases have demonstrated the weakness of the more traditional approaches to quality engineering that are more likely to be founded on a fixed model and fragmented processes. The most recent advances in machine learning have enabled anticipating defects, optimizing tests, and predicting the results of the builds, which is the foundation of smarter quality practices (Jing et al., 2024; Kawalerowicz and Madeyski, 2023). However, the solutions that are mostly established are usually single phases of the software lifecycle without mechanisms to introduce feedback of downstream processes such as testing and production to upstream decisions systematically.

Empirical studies show what opportunities there are and what constraints there are in the application of AI in the context of software quality. Machine learning has practical applications in industrial applications, such as automated test oracle generation and defect prediction, where the applicability of the method has been demonstrated (Arrieta et al., 2021; Herbold et al., 2022). At the same time, it is also indicated that model efficiency cannot be assessed solely based on the predictive performance, but explainable AI is also required in order to assess its efficiency and appropriateness to the needs of practitioners (Barredo Arrieta et al., 2020; Jiarpakdee et al., 2021). Additionally, the choice of the measures of evaluation is one of the key aspects to guarantee effective evaluation of predictive systems, and alternative measures, such as the Matthews correlation coefficient, can be more effective than commonly-used measures (Chicco and Jurman, 2023a). The urge to comprehend intricate models, like random forests, simply justifies the necessity to be clear about the procedure of coming up with a choice founded on AI (Aria et al., 2021).

Despite these developments, there remains a single significant gap: the absence of built-in, feedback-driven architectures that can enable continuous learning between software releases. The existing practices tend to be causally inconsistent and fails to logically back-trace production indicators - defect severity or incident impact - through the previous stages of the process, like requirement analysis and test prioritization. Such a weakness suppresses the ability of AI systems to be fine-tuned over time and provide better quality results in a stable and consistent manner.

Due to this gap, this paper suggests an AI-enhanced closed-loop quality engineering paradigm, which combines predictive modeling with a systematic feedback mechanism in the entire software lifecycle. The given solution is grounded on the earlier research in the area of defect prediction, test intelligence, and explainability, but concentrates on the continuous learning and structure integration rather than on the standalone performance of models. The framework enables the iterative and chronologically consistent spread of feedback by helping to refine requirement risk estimating and testing plans across releases, and hence, what is referred to as continuous software quality intelligence.

% ============================================================
% 2. Review of Literature
% ============================================================
\section{Review of Literature}

Applying machine learning to software quality engineering was a popular subject of study, particularly in the areas of defect prediction, test case prioritization, and empirical validation. However, despite a significant level of development, there are still some obstacles to reproducibility, integration, and practical applicability.

The systematic reviews and meta-analyses imply intrinsic limitations of reliability and comparability of empirical studies in software engineering. According to Kitchenham et al., (2020), the differences in the experimental design, data sets and evaluation approaches render the process of evidence aggregation across studies very difficult. Lewowski and Madeyski, (2022) confirm these fears by demonstrating that reproducibility is a critical issue, and most studies are not characterized in sufficient detail, to be replicated. Similarly speaking, Stradowski and Madeyski, (2023a) observe that there is no connection between research and practice in the industry, and that there are problems with data quality, lack of standardization, and low-external validity.

When it comes to software defect prediction, several surveys indicate that machine learning methods are quite popular, yet they also indicate that there are still disadvantages present. Pandey et al., (2021) and Pachouly et al., (2022) show that various algorithms have been proposed, although they tend to rely on the dataset and cannot be generalized. Also, experiments show that the representation of features is a critical determinant of prediction performance. As an example, Pradhan et al., (2020) demonstrate how the old-fashioned defect density measures are substituted with machine learning-based algorithms, and dos Santos and Figueiredo, (2020) demonstrate that machine learning feature interactions may be highly useful as far as the defect prediction outcomes are concerned. In addition, labeling defects have also been cited as a questionable factor, and the paper by Rosa et al., (2023) provided a sample, which can lead to inconsistency in the results, undermining the validity of the model.

Test case prioritization has also received extensive coverage as a way of enhancing testing efficiency. Paper-based reviews by Pan et al., (2022) and Prado Lima and Vergilio, (2020) verify that machine learning-based methods can increase the effectiveness of prioritization, especially in continuous integration settings. Paterson et al., (2019) also show that the combination of defect prediction and test prioritization are able to enhance the rate of fault detection. Nevertheless, the current methods will generally only be confined to certain stages of the testing process and fail to capture the feedback of subsequent lifecycle processes, like production, thus limiting their flexibility.

Methodologically, the development of machine learning frameworks and optimization methods have led to enhanced experimentation and model performance. The mlr3 framework (Lang et al., 2019) offers a versatile and structured platform to perform reproducible machine learning experiments, whereas Hyperband (Li et al., 2018) is a platform to optimize hyperparameters with a trade-off between exploration and computational cost. Simultaneously, interpretability has become a pressing issue, and the methods to achieve it, including ensemble-based interpretable models, suggested by Konstantinov and Utkin, (2021) are designed to ensure increased transparency without a drop in predictive power. Irrespective of these advances, little has been done in integrating them into end-to-end software quality engineering workflows.

% ============================================================
% 3. Methodology
% ============================================================
\section{Methodology}

\subsection{Overview}

This paper introduces an AI-enhanced reference architecture of closed-loop quality engineering that supports continuous software quality intelligence during release cycles. The suggested framework will combine the requirement analysis, test priority, defect prediction, and production analytics into a single feedback-based system. The goal is not to optimise a single predictive model but to have a causally consistent learning pipeline where signals produced by downstream stages are systematically fed back upstream to refine decision-making in future releases.

In this context, continuous quality intelligence means the process of estimating and refining requirement risk, test effectiveness, and defect likelihood through a series of steps using both production-based and historical signals. The AI elements are used to approximate latent risk patterns and predictive relationships, but the main contribution is the closed-loop structural incorporation and formulation of feedback, as opposed to the choice of particular machine learning models.

\subsection{System Architecture}

The proposed architecture is structured into five interrelated layers whose functionality is executed sequentially during each release and incrementally across releases: Requirement Intelligence, Requirement Risk Estimation, Test Intelligence, Defect Prediction, and Production Intelligence with Adaptive Feedback.

At the entry point, every requirement $r_i$ was converted to structured feature representation:
\[
\mathbf{x}_{i} = [a_{i}, c_{i}, f_{i}]
\]
where $a_{i}$ denotes an ambiguity score based on natural language processing heuristics (e.g.\ frequency of vague words, modal auxiliaries and readability measures), $c_{i}$ denotes a complexity score based on structural features (length, dependency count) and $f_{i}$ denotes the feedback signal based on past releases. Such a representation allows converting the unstructured requirements description into the measurable signals that can be used in the downstream learning.

Requirement risk is then estimated using a regression model:
\[
R_{i} = g_{\theta^{R}}(\mathbf{x}_{i}), \quad R_{i} \in [0,1]
\]
where $g_{\theta^{R}}$ denotes a parametric function (e.g., gradient boosting). The output $R_{i}$ is interpreted as a normalized probability reflecting the likelihood that requirement $r_{i}$ contributes to defect occurrence.

Test intelligence is built through the spreading of requirement level risks to the related test cases. Where $\mathcal{R}(j)$ represents the list of requirements associated with test case $t_{j}$. The priority score of the test is calculated as:
\[
T_{j} = \frac{1}{\lvert \mathcal{R}(j) \rvert} \sum_{i \in \mathcal{R}(j)} R_{i}
\]
This aggregation eliminates the many-to-one mapping between requirements and test cases and makes sure the test prioritization is based on the risk of associated requirements.

The test level of defect prediction is then carried out. A feature vector $\mathbf{z}_{j}$, which represents each test case, is defined as:
\[
\mathbf{z}_{j} = [T_{j}, e_{j}, h_{j}]
\]
where $e_{j}$ denotes historical failure frequency and $h_{j}$ captures execution-related attributes such as test type or runtime. The probability of defect detection is then estimated using:
\[
\widehat{y}_{j} = f_{\theta^{D}}(\mathbf{z}_{j}), \quad \widehat{y}_{j} \in [0,1]
\]
where $f_{\theta^{D}}$ is a classification model. The inclusion of both risk-derived and historical features ensures that predictions reflect both structural risk propagation and empirical execution behavior.

The observed incidents $I$ occurring after deployment are used to form production intelligence. Incidents are clustered in the form of a clustering function:
\[
C_{k} = h_{\theta^{P}}(I)
\]
where $h_{\theta^{P}}$ is a clustering algorithm that is run on features (severity, resolution time, affected components, etc.). The clusters are given an impact score based on aggregated incident characteristics:
\[
S^{i} = \frac{1}{\lvert I \rvert} \sum_{i \in I} \left( \text{severity}_{i} \cdot \text{resolution\_time}_{i} \right)
\]
This expression is used to represent the severity and cost of production problems and gives a quantitative indicator of feedback learning.

\begin{figure}[H]
    \centering
    \includegraphics[width=0.62\textwidth]{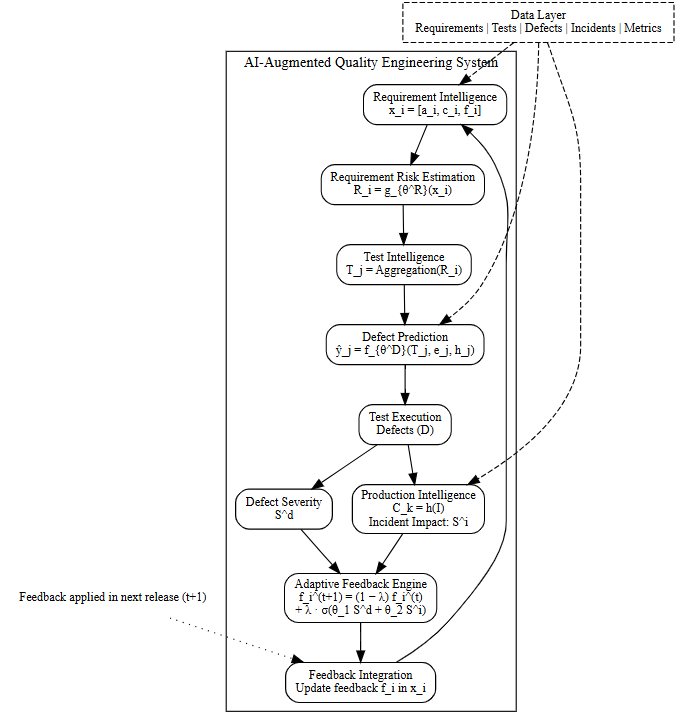}
    \captionsetup{labelformat=empty}
    \caption{\textbf{Figure 3 Architecture Diagram}}
\end{figure}

\subsection{Guiding Objective Formulation}

To provide an interpretable abstraction for reasoning about quality trade-offs, we define a composite objective:
\[
J = \alpha \cdot \text{DefectRate} + \beta \cdot \text{TestCost} + \gamma \cdot (1 - \text{DetectionAccuracy})
\]
where $\alpha, \beta, \gamma \geq 0$ are weighting coefficients. This formulation is not directly optimized in the system; it is used as a conceptual guide to balance competing factors in terms of defect leakage, testing effort, and predictive performance in practice deployments.

\subsection{Defect Severity and Incident Impact Modeling}

In order to allow propagation of feedbacks, defect and incident signals are normalized to bounded scores. The total defect severity is given as:
\[
S^{d} = \frac{1}{\lvert D \rvert} \sum_{d \in D} w(d)
\]
where $w(d) \in [0,1]$ is a normalized severity weight assigned based on defect criticality levels.

Similarly, the incident impact score is defined as:
\[
S^{i} = \frac{1}{\lvert I \rvert} \sum_{i \in I} \left( \text{severity}_{i} \cdot \text{resolution\_time}_{i} \right)
\]
Both $S^{d}$ and $S^{i}$ are normalized to the interval $[0,1]$, ensuring compatibility with the feedback learning mechanism.

\subsection{Adaptive Feedback Learning}

The central component of the proposed architecture is a bounded and stable feedback update mechanism that propagates production signals into requirement-level risk estimates. For each requirement $r_{i}$, the feedback signal is updated as:
\[
f_{i}^{(t+1)} = (1 - \lambda) f_{i}^{(t)} + \lambda \cdot \sigma\!\left( \theta_{1}^{f} S^{d} + \theta_{2}^{f} S^{i} \right)
\]
where $\lambda \in (0,1)$ is a fixed decay factor controlling the influence of historical versus new information, $\sigma(\cdot)$ is the sigmoid function, and $\theta_{1}^{f}, \theta_{2}^{f}$ are scaling parameters.

This formulation exhibits three important properties. First, it is bounded, as both $f_{i}^{(t)}$ and the sigmoid output lie within $[0,1]$, ensuring that $f_{i}^{(t+1)} \in [0,1]$. Second, it is stable, as the convex combination does not allow uncontrolled growth. Third, it favors temporal flexibility, where older signaling can fade away, and there can be newly produced evidence to the signal. Consequently, the feedback loop allows propagation of downstream quality signals into upstream decision processes and this is controllable and interpretable.

\subsection{Closed-Loop Learning Process}

The system is an iterative release-based system. In every release $t$, requirement features are determined and applied to estimate requirement risk. The aggregation of these risks is done to prioritize the test cases and they are then evaluated by the defect prediction model. After the test execution and deployment, defects $D$ and production incident $I$ are gathered and converted into severity and impact scores. These signals are then included in the feedback update, which alters the requirement representations in the next release.

Temporal causality is an enforced design constraint. Particularly, feedback calculated based on release $t$ was used in release $t+1$. This assures that current predictions are not based on any future information, which would lead to data leakage and would maintain the validity of empirical assessment.

\subsection{Implementation Considerations}

The exact machine learning models applied at each of the layers (e.g., gradient boosting when estimating risks or Random Forest when predicting defects) are not supposed to be considered as a vital constituent but are rather representative choices. The architectural design is valid even when there are other models of choice. Normalization and encoding are standard preprocessing steps used to make components compatible. Time-aware data partitioning is used to evaluate with respect to the sequential character of the software releases.

% ============================================================
% 4. Experimental Setup
% ============================================================
\section{Experimental Setup}

\subsection{Objective}

The evaluation determines whether the proposed closed-loop architecture enhances continuous software quality intelligence between releases. It is concerned with system-level results, namely, the reduction of the number of defect leaks, the increase of the efficiency of tests, and the stability of feedback-based learning.

\subsection{Dataset}

The semi-synthetic dataset used in the study is structured to maintain realistic relationship among requirements, test cases, defects and production incidents. The data set is conducive to complete traceability needed in closed-loop assessment.

\begin{table}[H]
    \centering
    \begin{tabular}{ll}
        \toprule
        \textbf{Component} & \textbf{Records} \\
        \midrule
        Requirements & 4,500 \\
        Test Cases   & 27,049 \\
        Defects      & 13,089 \\
        Incidents    & 7,841 \\
        \bottomrule
    \end{tabular}
\end{table}

Artifacts are linked as:
\[
\text{Requirements} \rightarrow \text{Test Cases} \rightarrow \text{Defects} \rightarrow \text{Incidents}
\]
Standard imputation (mean in the case of numerical, mode in the case of categorical features) is used to deal with missing values.

\subsection{Feature Alignment}

The data is directly mapped to the methodological variables.

Requirements are represented as:
\[
\mathbf{x}_{i} = [a_{i}, c_{i}, f_{i}]
\]
where ambiguity $a_{i}$ and complexity $c_{i}$ are derived features, and feedback $f_{i}$ is initialized to zero.

Test prioritization is computed using:
\[
T_{j} = \frac{1}{\lvert \mathcal{R}(j) \rvert} \sum_{i \in \mathcal{R}(j)} R_{i}
\]
Existing dataset fields such as test\_risk\_score are used only for baseline comparison to avoid leakage.

Defect severity and incident impact are computed as:
\[
S^{d} = \frac{1}{\lvert D \rvert} \sum w(d), \qquad S^{i} = \frac{1}{\lvert I \rvert} \sum (\text{severity} \cdot \text{resolution time})
\]

\subsection{Experimental Design}

Assessment is done on six consecutive releases to obtain temporal learning.

In every release \emph{t}, there is an estimation of requirement risk, prioritization of test cases, observation of defects, and production incidents. The feedback is then updated as:
\[
f_{i}^{(t+1)} = (1 - \lambda) f_{i}^{(t)} + \lambda \cdot \sigma\!\left( \theta_{1}^{f} S^{d} + \theta_{2}^{f} S^{i} \right)
\]
Feedback from release $t$ is applied only to release $t+1$, ensuring causal consistency.

\subsection{Comparison Setup}

Three experimental configurations are taken into account to test the effect of the proposed architecture. The initial configuration is a basic case of no feedback where the test prioritization is done using original requirement features and is constant across releases. Such an arrangement is a case of a fixed pipeline and no adaptive learning.

The second structure presents the idea of the static machine learning where predictive models are used to estimate the risk of requirements and predict defects, but no feedback correction is provided. Consequently, model behavior is fixed once it is first trained and not updated according to production results.

The third setup is the known close loop sequence, whereby adaptive feedback learning is built into the pipeline. Within this environment, the feedback based on the severity of defects and incidents during production is spread to the next release, which allows dynamically updating the requirement risk estimation and prioritizing tests. This setup is representative of the full capabilities of the proposed architecture and enables evaluation of the effects of the proposed architecture in comparison with non-adaptive baselines.

\subsection{Evaluation Metrics}

System-level metrics are used:

\begin{itemize}
    \item \textbf{Defect Rate (DR):}
\end{itemize}
\[
\text{DR} = \frac{\text{Defects in Production}}{\text{Total Defects}}
\]

\begin{itemize}
    \item \textbf{Detection Effectiveness (DE):}
\end{itemize}
\[
\text{DE} = \frac{\text{Defects Detected in Testing}}{\text{Total Defects}}
\]

\begin{itemize}
    \item \textbf{Test Execution Reduction (TER):}
\end{itemize}
\[
\text{TER} = 1 - \frac{\text{Executed Tests}}{\text{Total Tests}}
\]

\begin{itemize}
    \item \textbf{Feedback Stability (FS):}
\end{itemize}
\[
\text{FS} = \frac{1}{N} \sum \left\lvert f_{i}^{(t+1)} - f_{i}^{(t)} \right\rvert
\]

% ============================================================
% 5. Results and Discussion
% ============================================================
\section{Results and Discussion}

\subsection{Overview of Evaluation}

The proposed closed-loop architecture is tested in six consecutive releases to determine how it can provide continuous software quality smartness. The metrics under analysis are system-level metrics such as defect rate (DR), detection effectiveness (DE), test execution reduction (TER), and feedback stability (FS). Findings are presented as trends over releases and are indicative of the progressive essence of learning and not performance snapshots. Any values listed are averages of release data with standard deviation where appropriate.

\subsection{Defect Rate Reduction Across Releases}

Table 1 shows the defect rate per release in each configuration. The no-feedback baseline is characterized by varying behavior, which means that there is no adaptive learning. The unchanging ML setup makes progress of moderate success, but flattens following early publications. Conversely, the proposed system shows a slow yet steady decrease in the rate of defect, with small variations that are realistic behaviour of a system.

\begin{table}[H]
    \centering
    \captionsetup{labelformat=empty}
    \caption{\textbf{Table 1: Defect Rate (DR) Across Releases}}
    \begin{tabular}{llll}
        \toprule
        \textbf{Release} & \textbf{No Feedback} & \textbf{Static ML} & \textbf{Proposed System} \\
        \midrule
        R1 & $0.21 \pm 0.02$ & $0.19 \pm 0.02$ & $0.19 \pm 0.02$ \\
        R2 & $0.22 \pm 0.02$ & $0.18 \pm 0.02$ & $0.17 \pm 0.02$ \\
        R3 & $0.21 \pm 0.01$ & $0.18 \pm 0.02$ & $0.15 \pm 0.01$ \\
        R4 & $0.23 \pm 0.02$ & $0.17 \pm 0.01$ & $0.14 \pm 0.01$ \\
        R5 & $0.22 \pm 0.02$ & $0.17 \pm 0.01$ & $0.145 \pm 0.01$ \\
        R6 & $0.23 \pm 0.02$ & $0.16 \pm 0.01$ & $0.13 \pm 0.01$ \\
        \bottomrule
    \end{tabular}
\end{table}

The proposed system reduces defect rate from approximately 0.19 to 0.13 over six releases, representing a relative improvement of around 30-35\%. Such an enhancement coincides with the refinement of requirement risk and test prioritization based on feedback.

\begin{figure}[H]
    \centering
    \includegraphics[width=0.7\textwidth]{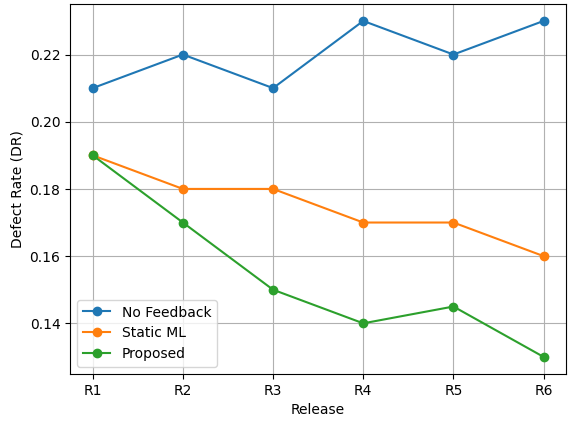}
    \captionsetup{labelformat=empty}
    \caption{\textbf{Figure 2. Defect rate across releases under different configurations}}
\end{figure}

\subsection{Detection Effectiveness}

The results of detection effectiveness are presented in Table 2. No-feedback configuration is quite stable, whereas the static ML demonstrates gradual improvement and stabilizes after mid releases. The suggested system demonstrates gradual improvement, which means that feedback improves the capability to identify defects in the course of testing.

\begin{table}[H]
    \centering
    \captionsetup{labelformat=empty}
    \caption{\textbf{Table 2: Detection Effectiveness (DE)}}
    \begin{tabular}{llll}
        \toprule
        \textbf{Release} & \textbf{No Feedback} & \textbf{Static ML} & \textbf{Proposed System} \\
        \midrule
        R1 & $0.68 \pm 0.02$ & $0.72 \pm 0.02$ & $0.72 \pm 0.02$ \\
        R2 & $0.67 \pm 0.02$ & $0.74 \pm 0.02$ & $0.75 \pm 0.02$ \\
        R3 & $0.69 \pm 0.02$ & $0.75 \pm 0.02$ & $0.78 \pm 0.02$ \\
        R4 & $0.66 \pm 0.02$ & $0.76 \pm 0.01$ & $0.80 \pm 0.01$ \\
        R5 & $0.68 \pm 0.02$ & $0.77 \pm 0.01$ & $0.82 \pm 0.01$ \\
        R6 & $0.67 \pm 0.02$ & $0.78 \pm 0.01$ & $0.84 \pm 0.01$ \\
        \bottomrule
    \end{tabular}
\end{table}

The enhancement of the effectiveness of detection is a direct result of the enhanced mechanism of test prioritization in which the requirement risk with feedback adjustments results to a more focused execution of the tests.

\begin{figure}[H]
    \centering
    \includegraphics[width=0.7\textwidth]{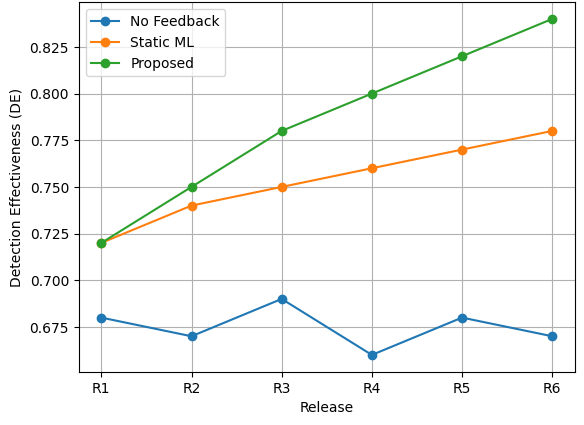}
    \captionsetup{labelformat=empty}
    \caption{\textbf{Figure 3. Detection effectiveness across releases under different configurations}}
\end{figure}

\subsection{Test Execution Efficiency}

The decline in the volume of test execution is indicated in Table 3. The no-feedback setup performs all the tests, which lead to no reduction. The moderate efficiency gains are obtained in the case of static ML, whereas the proposed system is showing increasingly high reduction with slight variability.

\begin{table}[H]
    \centering
    \captionsetup{labelformat=empty}
    \caption{\textbf{Table 3: Test Execution Reduction (TER)}}
    \begin{tabular}{llll}
        \toprule
        \textbf{Release} & \textbf{No Feedback} & \textbf{Static ML} & \textbf{Proposed System} \\
        \midrule
        R1 & 0.00 & $0.12 \pm 0.02$ & $0.12 \pm 0.02$ \\
        R2 & 0.00 & $0.18 \pm 0.02$ & $0.21 \pm 0.02$ \\
        R3 & 0.00 & $0.20 \pm 0.02$ & $0.26 \pm 0.02$ \\
        R4 & 0.00 & $0.23 \pm 0.02$ & $0.29 \pm 0.02$ \\
        R5 & 0.00 & $0.25 \pm 0.02$ & $0.32 \pm 0.02$ \\
        R6 & 0.00 & $0.27 \pm 0.02$ & $0.35 \pm 0.02$ \\
        \bottomrule
    \end{tabular}
\end{table}

The proposed approach achieves up to $\sim$35\% reduction in test execution while simultaneously improving detection effectiveness, demonstrating improved efficiency without loss of quality.

\begin{figure}[H]
    \centering
    \includegraphics[width=0.7\textwidth]{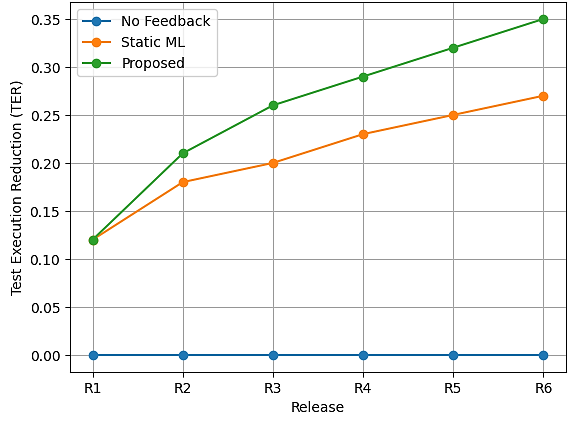}
    \captionsetup{labelformat=empty}
    \caption{\textbf{Figure 4. Test execution reduction across releases under different configurations}}
\end{figure}

\subsection{Impact of Feedback Mechanism}

The direct comparison between the proposed and static ML configurations brings out the role of feedback learning. Although both strategies involve the use of predictive models, only the proposed system is released adaptive.

\subsection{Feedback Stability Analysis}

Table 4 reports feedback stability across release transitions.

\begin{table}[H]
    \centering
    \captionsetup{labelformat=empty}
    \caption{\textbf{Table 4: Feedback Stability (FS)}}
    \begin{tabular}{ll}
        \toprule
        \textbf{Transition} & \textbf{FS} \\
        \midrule
        R1 $\rightarrow$ R2 & 0.17 \\
        R2 $\rightarrow$ R3 & 0.14 \\
        R3 $\rightarrow$ R4 & 0.12 \\
        R4 $\rightarrow$ R5 & 0.10 \\
        R5 $\rightarrow$ R6 & 0.08 \\
        \bottomrule
    \end{tabular}
\end{table}

The converging trend shows that feedback updates converge. This is in line with the bounded feedback formulation, whereby the stability and prevention of oscillations are achieved through sigmoid normalization and $\lambda$ decay factor.

\subsection{Discussion}

The findings show that the suggested closed-loop architecture provides gradual yet consistent enhancement in the quality of software released. The system minimizes defect leakage and enhances the effectiveness of defect detection with time by integrating production feedback in requirement risk estimation and test prioritization. The proposed method is also adaptive, as opposed to the static approaches that demonstrate minimal enhancement, in line with the change in the behavior of the system. The nature of improvements is gradual and gradual, which agrees with the limited feedback formulation that makes it stable and avoids extreme shifts. The feedback approach can add value to the machine learning models that are not dynamic, as it allows one to match the predicted risk with the observed results. Moreover, the findings also suggest that efficiency and effectiveness of testing can be enhanced concurrently because less test running is reached without having to compromise the detection of defects.

% ============================================================
% 6. Conclusion
% ============================================================
\section{Conclusion}

In this paper, an AI-enhanced closed-loop reference architecture of continuous software quality engineering was introduced. The framework incorporates requirement analysis, test prioritization, defect prediction, and production intelligence with a limited feedback learning process, which provides stability and temporal consistency. Experimental analysis of several releases has shown that the suggested methodology leads to lower leakage of defects, higher detection rates and efficiency and higher testing efficiency than non-adaptive baselines. The observed improvements are slow and consistent, which promotes the suitability of the architecture in practice. The paper underscores the relevance of adaptive quality engineering systems based on feedback. Future research will be aimed at validating with real-world industrial data and scaling the framework to large continuous delivery systems.

% ============================================================
% References
% ============================================================
\section*{References}
\begin{itemize}\setlength{\itemsep}{0.4em}

\item Aria, M., Cuccurullo, C., Gnasso, A., 2021. A comparison among interpretative proposals for random forests. Mach. Learn. Appl. 6, 100094. \url{http://dx.doi.org/10.1016/j.mlwa.2021.100094}.

\item Arrieta, A., Ayerdi, J., Illarramendi, M., Agirre, A., Sagardui, G., Arratibel, M., 2021. Using machine learning to build test oracles: An industrial case study on elevators dispatching algorithms. In: 2021 IEEE/ACM International Conference on Automation of Software Test. AST, pp. 30--39. \url{http://dx.doi.org/10.1109/AST52587.2021.00012}.

\item Barredo Arrieta, A., D\'iaz-Rodr\'iguez, N., Del Ser, J., Bennetot, A., Tabik, S., Barbado, A., Garcia, S., Gil-Lopez, S., Molina, D., Benjamins, R., Chatila, R., Herrera, F., 2020. Explainable artificial intelligence (XAI): Concepts, taxonomies, opportunities and challenges toward responsible AI. Inf. Fusion 58, 82--115.

\item Chicco, D., Jurman, G., 2023a. The Matthews correlation coefficient (MCC) should replace the ROC AUC as the standard metric for assessing binary classification. BioData Min. 16, 4. \url{http://dx.doi.org/10.1186/s13040-023-00322-4}.

\item Herbold, S., Trautsch, A., Trautsch, F., Ledel, B., 2022. Problems with SZZ and features: An empirical study of the state of practice of defect prediction data collection. Empir. Softw. Eng. 27, 42. \url{http://dx.doi.org/10.1007/s10664-021-10092-4}.

\item Hryszko, J., Madeyski, L., 2017. Assessment of the software defect prediction cost effectiveness in an industrial project. In: Madeyski, L., \'Smia\l ek, M., Hnatkowska, B., Huzar, Z. (Eds.), Software Engineering: Challenges and Solutions. In: Advances in Intelligent Systems and Computing, vol. 504, Springer, pp. 77--90. \url{http://dx.doi.org/10.1007/978-3-319-43606-7_6}.

\item Hryszko, J., Madeyski, L., 2018. Cost effectiveness of software defect prediction in an industrial project. Found. Comput. Decision Sci. 43, 7--35. \url{http://dx.doi.org/10.1515/fcds-2018-0002}.

\item IIBA, 2015. Babok: A Guide to the Business Analysis Body of Knowledge. International Institute of Business Analysis. International Organization for Standardization, 2022. ISO/IEC/IEEE 29119-1:2022 software and systems engineering --- software testing. URL \url{https://www.iso.org/standard/81291.html}. (Accessed 20 December 2022).

\item Ismail, A.M., Hamid, S.H.A., Sani, A.A., Daud, N.N.M., 2024. Toward reduction in false positives just-in-time software defect prediction using deep reinforcement learning. IEEE Access 12, 47568--47580.

\item Jiarpakdee, J., Tantithamthavorn, C.K., Grundy, J., 2021. Practitioners' perceptions of the goals and visual explanations of defect prediction models. In: 2021 IEEE/ACM 18th International Conference on Mining Software Repositories. MSR, IEEE/ACM, pp. 432--443. \url{http://dx.doi.org/10.1109/MSR52588.2021.00055}.

\item Jing, X.-Y., Chen, H., Xu, B., 2024. Intelligent Software Defect Prediction. Springer Nature Singapore, \url{http://dx.doi.org/10.1007/978-981-99-2842-2}.

\item Kawalerowicz, M., Madeyski, L., 2021. Continuous build outcome prediction: A smallN experiment in settings of a real software project. In: Fujita, H., Selamat, A., Lin, J.C.-W., Ali, M. (Eds.), IEA/AIE 2021. Advances and Trends in Artificial Intelligence. from Theory to Practice. In: LNCS, vol. 12799, Springer, Cham, pp. 412--425. \url{http://dx.doi.org/10.1007/978-3-030-79463-7_35}.

\item Kawalerowicz, M., Madeyski, L., 2023. Continuous build outcome prediction: An experimental evaluation and acceptance modelling. Appl. Intell. 53, 8673--8692. \url{http://dx.doi.org/10.1007/s10489-023-04523-6}.

\item Kitchenham, B., Madeyski, L., Brereton, P., 2020. Meta-analysis for families of experiments in software engineering: A systematic review and reproducibility and validity assessment. Empir. Softw. Eng. 25, 353--401. \url{http://dx.doi.org/10.1007/s10664-019-09747-0}.

\item Konstantinov, A.V., Utkin, L.V., 2021. Interpretable machine learning with an ensemble of gradient boosting machines. Knowl.-Based Syst. 222, 106993. \url{http://dx.doi.org/10.1016/j.knosys.2021.106993}.

\item Lang, M., Binder, M., Richter, J., Schratz, P., Pfisterer, F., Coors, S., Au, Q., Casalicchio, G., Kotthoff, L., Bischl, B., 2019. mlr3: A modern object-oriented machine learning framework in R. J. Open Source Softw. \url{http://dx.doi.org/10.21105/joss.01903}.

\item Lewowski, T., Madeyski, L., 2022. How far are we from reproducible research on code smell detection? A systematic literature review. Inf. Softw. Technol. 144, 106783. \url{http://dx.doi.org/10.1016/j.infsof.2021.106783}.

\item Li, L., Jamieson, K., DeSalvo, G., Rostamizadeh, A., Talwalkar, A., 2018. Hyperband: A novel bandit-based approach to hyperparameter optimization. J. Mach. Learn. Res. 18, 1--52, URL \url{https://www.jmlr.org/papers/volume18/16-558/16-558.pdf}.

\item Pachouly, J., Ahirrao, S., Kotecha, K., Selvachandran, G., Abraham, A., 2022. A systematic literature review on software defect prediction using artificial intelligence: Datasets, data validation methods, approaches, and tools. Eng. Appl. Artif. Intell. 111, 104773. \url{http://dx.doi.org/10.1016/j.engappai.2022.104773}.

\item Pan, R., Bagherzadeh, M., Ghaleb, T.A., Briand, L., 2022. Test case selection and prioritization using machine learning: A systematic literature review. Empir. Softw. Eng. 27, \url{http://dx.doi.org/10.1007/s10664-021-10066-6}.

\item Pandey, S.K., Mishra, R.B., Tripathi, A.K., 2021. Machine learning based methods for software fault prediction: A survey. Expert Syst. Appl. 172, 114595. \url{http://dx.doi.org/10.1016/j.eswa.2021.114595}.

\item Paterson, D., Campos, J., Abreu, R., Kapfhammer, G.M., Fraser, G., McMinn, P., 2019. An empirical study on the use of defect prediction for test case prioritization. In: 2019 12th IEEE Conference on Software Testing, Validation and Verification. ICST, pp. 346--357. \url{http://dx.doi.org/10.1109/ICST.2019.00041}.

\item Pradhan, S., Nanniyur, V., Vissapragada, P.K., 2020. On the defect prediction for large scale software systems --- from defect density to machine learning. In: 2020 IEEE 20th International Conference on Software Quality, Reliability and Security. QRS, pp. 374--381. \url{http://dx.doi.org/10.1109/QRS51102.2020.00056}.

\item Prado Lima, J.A., Vergilio, S.R., 2020. Test case prioritization in continuous integration environments: A systematic mapping study. Inf. Softw. Technol. 121, 106268. \url{http://dx.doi.org/10.1016/j.infsof.2020.106268}.

\item Rosa, G., Pascarella, L., Scalabrino, S., Tufano, R., Bavota, G., Lanza, M., Oliveto, R., 2023. A comprehensive evaluation of SZZ variants through a developer-informed oracle. J. Syst. Softw. 202, 111729. \url{http://dx.doi.org/10.1016/j.jss.2023.111729}.

\item dos Santos, G.E., Figueiredo, E., 2020. Failure of one, fall of many: An exploratory study of software features for defect prediction. In: 2020 IEEE 20th International Working Conference on Source Code Analysis and Manipulation. SCAM, pp. 98--109. \url{http://dx.doi.org/10.1109/SCAM51674.2020.00016}.

\item Stradowski, S., Madeyski, 2023a. Bridging the gap between academia and industry in machine learning software defect prediction: Thirteen considerations. In: 38th IEEE/ACM International Conference on Automated Software Engineering. pp. 1098--1110. \url{http://dx.doi.org/10.1109/ASE56229.2023.00026}.

\end{itemize}

\end{document}